\documentclass[twocolumn,aps,prb,showpacs,amsmath,superscriptaddress,longbibliography,notitlepage]{revtex4-1}
\usepackage{amssymb}
\usepackage{mathrsfs}
\usepackage{graphicx}
\usepackage{float}
\usepackage[caption=false]{subfig}
\usepackage[pdftex,colorlinks=red]{hyperref}
\usepackage{makecell}

\usepackage{epstopdf}
\usepackage[flushleft]{threeparttable}
\usepackage{booktabs}

\usepackage[normalem]{ulem}
\usepackage{verbatim}
\usepackage{xcolor}
\usepackage{bm}

\bibpunct{[}{]}{,}{n}{}{}

\def\blue{\textcolor{blue}}

\def\brown{\textcolor{brown}}

\def\lrj{\textcolor{purple}}

\begin{document}

\title{Topologically compatible non-Hermitian skin effect}
\author{Rijia Lin}
 \affiliation{Guangdong Provincial Key Laboratory of Quantum Metrology and Sensing $\&$ School of Physics and Astronomy, Sun Yat-Sen University (Zhuhai Campus), Zhuhai 519082, China}
\author{Linhu Li}  \thanks{corresponding author: lilh56@mail.sysu.edu.cn}
\affiliation{Guangdong Provincial Key Laboratory of Quantum Metrology and Sensing $\&$ School of Physics and Astronomy, Sun Yat-Sen University (Zhuhai Campus), Zhuhai 519082, China}

\begin{abstract}
  The bulk-boundary correspondence (BBC) relates 
  in-gap boundary modes to bulk topological invariants.
  In certain non-Hermitian topological systems, conventional BBC becomes invalid in the presence of the non-Hermitian skin effect (NHSE),
  which manifests as distinct
  energy spectra under the periodic and open boundary conditions
  and massive eigenstate localization at boundaries.
  In this work, we introduce a scheme to induce NHSE without breaking conventional BBC, dubbed as the topologically compatible NHSE (TC-NHSE).
  In a general one dimensional two-band model,  
  we unveil two types of TC-NHSE that do not alter topological phase transition points 
  under any circumstance or only in a certain parameter regime, respectively.
  Extending our model into two dimension,
  we find that TC-NHSE can be selectively compatible to different sets of Weyl points between different bands of the resultant semimetallic system, turning some of them into bulk Fermi arcs while keeping the rest unchanged.
  Our work hence helps clarify the intricate interplay between topology and NHSE in non-Hermitian systems, and provides a versatile approach for designing non-Hermitian topological systems where topological properties and NHSE do not interfere each other.
\end{abstract}

\date{\today}

\maketitle

\section{introduction}


The bulk-boundary correspondence (BBC) is among the most important principles in the investigation of topological phases of matter~\cite{hasan2010colloquiuma,qi2011topological,ryu2002topologicala,mong2011edge,schnyder2008classification,delplace2011zak}.
Namely, a topological invariant defined under the periodic boundary conditions (PBCs) predicts the number of topological boundary modes under the open boundary conditions (OBCs).
Accordingly, Topological phase transitions involve band gap closing for both PBC and OBC bulk spectra, which is the only possibility to change the topological invariant and the number of topological modes.

In recent years, investigations into non-Hermitian topological phases reveal an enigmatic violation of conventional BBC~\cite{lee2019anatomy,martinezalvarez2018nonhermitiana,lee2016anomalous,kunst2018biorthogonal,jin2019bulkboundary,edvardsson2019nonhermitiana}, as 
the eigenvalues and eigenvectors of most non-Hermitian systems are extremely sensitive to boundary conditions~\cite{xiong2018why,koch2020bulkboundary,edvardsson2022sensitivity}.
In particular, the topological structure of PBC energy bands can be changed by passing through exceptional points (EPs) when gradually eliminating the amplitude of hopping between the ends of a chain to reach OBCs, 
which leads to the connection of different bands in the complex energy plane~\cite{xiong2018why}.
In these systems, OBC eigenstates tend to be exponentially localized near system's boundaries, which is called the non-Hermitian skin effect (NHSE)~\cite{alvarez2018nonH,yao2018edge}, 
and the number of topological edge modes may no longer be related to the topological invariants formulated in terms of the Bloch Hamiltonian, as the gap closing condition can be different for PBCs and OBCs.
A modified BBC has been established for these systems through the non-Bloch band theory~\cite{yao2018edge,lee2019anatomy,yokomizo2019nonBloch,zhang2020correspondence}.
In such a framework, topological invariants are defined with complex deformations of lattice momentum, 
which generally have rather intricate explicit forms and extremely sensitive numerical solutions
for systems beyond minimal models~\cite{auxiliaryGBZ,lee2020unraveling}, making it difficult to accurately identify their topological properties.

  In this work, 
  we unveil a class of topological compatible NHSE (TC-NHSE), where massive edge localization of eigenstates emerges without breaking the conventional BBC, as non-Hermiticity is introduced properly to avoid inducing any EP to the system.
We first discuss the general scheme for inducing two types of TC-NHSE in two-band non-Hermitian Hamiltonians, 
where conventional BBC is preserved either universally or only within certain parameter regimes, respectively.
An explicit one-dimensional (1D) example of a non-Hermitian Su-Schrieffer-Heeger (SSH) model~\cite{su1979solitons} is then demonstrated accordingly.
We next extend the 1D SSH model into a 2D semimetallic system with an extra spin degree of freedom and spin-orbit coupling terms, where three sets of Weyl points protected by different (hidden) $\mathcal{PT}$-symmetries emerge between different energy bands.
More interestingly, TC-NHSE can be introduced in a way that is selectively compatible to different sets of Weyl points, turning some of them into EPs by breaking the corresponding symmetry while the others remain unchanged.
The topological compaticity of the induced NHSE is further verified by calculating the non-Hermitian Zak phase defined with Bloch eigenstates, which precisely predict topological properties under OBCs, for both the 1D and 2D models. 

\section{General two-band Hamiltonians}
We begin with a general non-Hermitian two-band Hamiltonian,
\begin{eqnarray}
h(k)=\sum_{\alpha=0,x,y,z}h_\alpha(k)\sigma_\alpha,\nonumber\\
h_{\alpha}=d_{\alpha}(k)+ig_{\alpha}(k),\label{eq:2band}
\end{eqnarray}
with $d(k)$ and $g(k)$ taking real values, describing the Hermitian and anti-Hermitian parts of the system, $\sigma_{x,y,z}$ the Pauli matrices and $\sigma_0$ the $2\times2$ identity matrix.
We will omit the $k$-dependency in the notations of $d_{\alpha}$, $g_{\alpha}$, and $h_{\alpha}$ for simplicity.
A topological phase transition requires a band touching point of $h(k)$, which occurs when
$$P(k)=\sum_{\alpha=x,y,z}h_\alpha^2=0,$$
namely
\begin{eqnarray}
\sum_{\alpha=x,y,z}(d_\alpha^2-g_\alpha^2)&=&0,~~\sum_{\alpha=x,y,z}d_\alpha g_\alpha=0,.
\end{eqnarray}
This condition generally applies to an exceptional degeneracy point (EP) of $h(k)$.
A normal degenerate point (DP) when the system further satisfies $h_\alpha=0$ for $\alpha=x,y,z$. 

The breakdown of conventional BBC requires different bands connect head to tail in the complex energy plane~\cite{xiong2018does,lee2019anatomya,liang2022topologicalb} , corresponding to a nontrivial winding of eigenvectors around an EP ~\cite{mailybaev2005geometric,yin2018geometricala,li2019geometric}.
As Hermitian systems forbidden any EP to exist, this means that if we start from a Hermitian $h(k)$ and turn on non-Hermiticity gradually, it must go through a EP transition to break conventional BBC. 
In turn, in order for NHSE to emerge without breaking the conventional BBC, we need to properly introduce non-Hermiticity to the system, and avoid inducing any EP. For our minimal 2-band Hamiltonian, this can be done by introducing
\begin{itemize}
\item[(i)] a nonzero $g_0$ that shifts both bands by the same amount for each momentum $k$, or
\item[(ii)]  nonzero $g_x$, $g_y$, or $g_z$, which vanishes together with the Hermitian parts of the Hamiltonian, i.e. when $d_x=d_y=d_z=0$.
\end{itemize}
In both cases, 
NHSE can emerge provided that the non-Hermitian terms also induce nontrivial spectral winding to the PBC spectrum.
Note that the first scheme shall avoid EP under any circumstance, as $g_0$ only shifts the eigenvalues of $h(k)$, but does not change its degenerate condition.
On the other hand, the second scheme 
does not always guarantee the absence of EP,
and the NHSE may be topologically compatible only in a certain parameter regime, as elaborated with the example model discussed below.


\section{TC-NHSE in a 1D two-band model} \label{sec:1d}
\begin{figure}
  \includegraphics[width=1\linewidth]{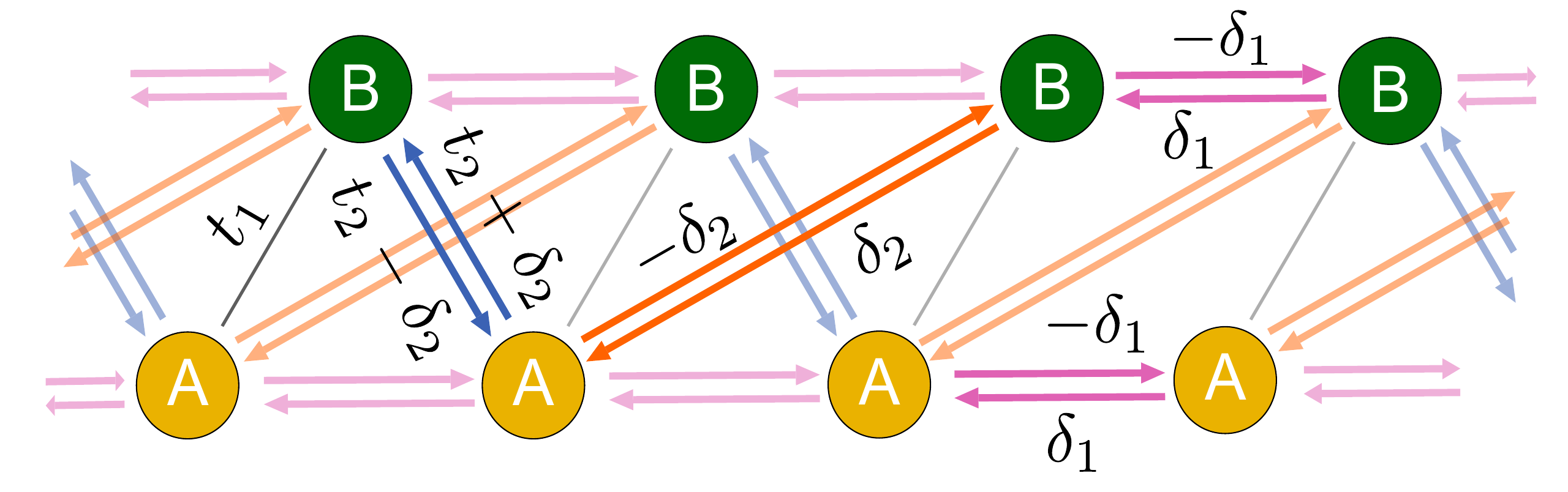}
  \caption{
  A sketch of the 1D lattice of the non-Hermitian SSH model described by Eqs. \eqref{eq:SSH} and \eqref{eq:nonH_SSH}. 
  }
  \label{fig:model}
  \end{figure}
\begin{figure}
  \includegraphics[width=1\linewidth]{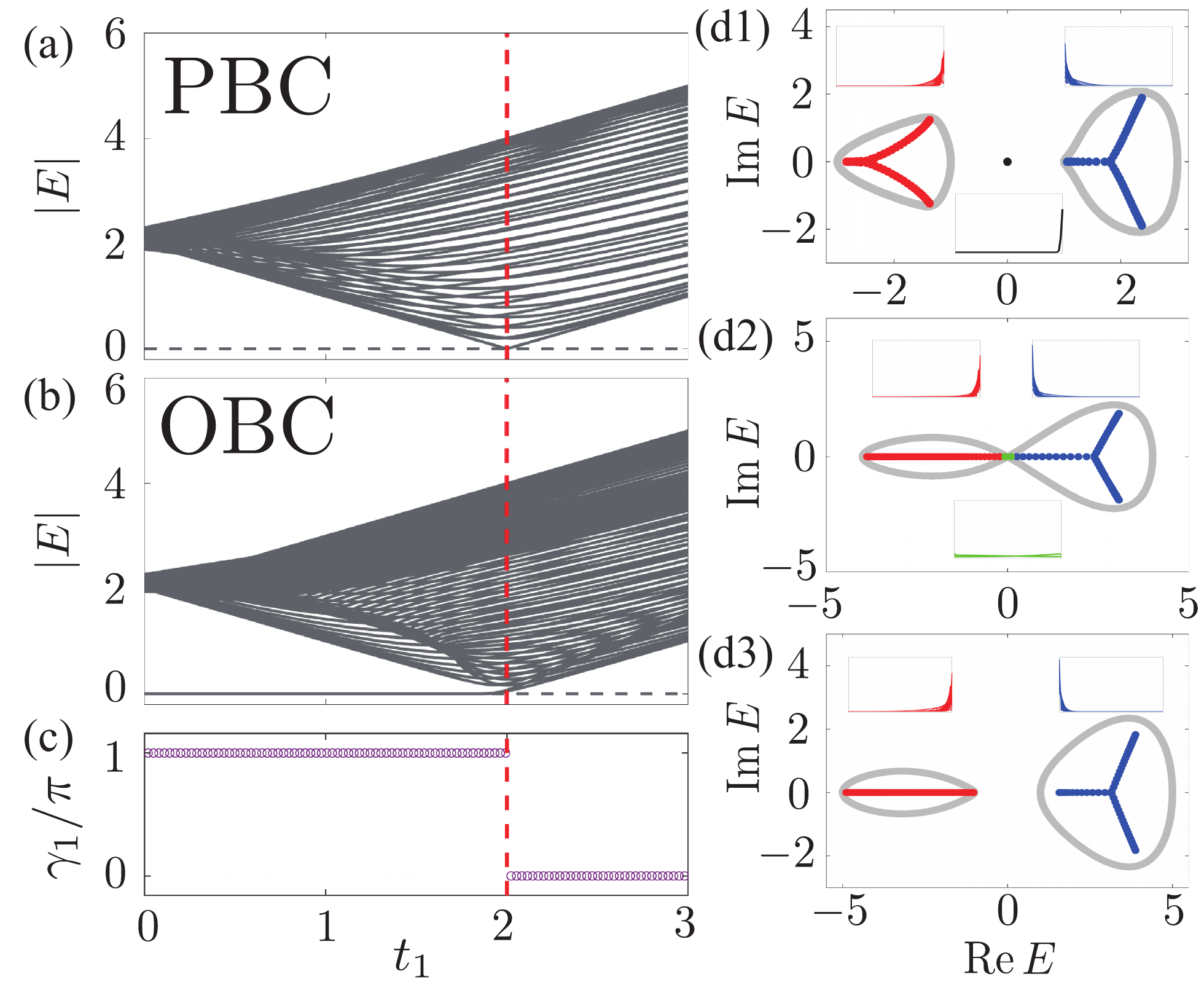}
  \caption{TC-NHSE in the non-Hermitian SSH model. 
  (a), (b) The absolute values of eigenenergies under PBCs and OBCs respectively, with the same gap-closing point at $t_1=2$ (red dash lines).
  (c) The Zak phase $\gamma_{1}$ as a function of $t_{1}$. 
  (d1) to (d3) The complex spectrum under the PBCs (gray) and OBCs (red, blue, black, and green) for $t_1=1$, $2$, and $3$ respectively, with insets demonstrating distribution profiles of OBC eigenstates with the same colors.
  Other parameters are $t_{2}=2$, $\delta_{1}=0.75$, $\delta_{2}=0.5$, $N=60$.
}
\label{fig:1d_spectrum}
\end{figure}

To give an explicit example, we consider a non-Hermitian extension of the Su-Schrieffer-Heeger (SSH) model~\cite{Su1979} as sketched in Fig. \ref{fig:model}, described by the Hamiltonian $H_{\rm 1D}=H_{\rm SSH}+H_{\rm nonH}$ where
\begin{eqnarray}
H_{\rm SSH}&=&\sum_n^N (t_1\hat{a}^\dagger_{n}\hat{b}_{n}+t_2\hat{b}^\dagger_{n}\hat{a}_{n+1})+h.c.,\label{eq:SSH}\\
H_{\rm nonH}&=&\sum_n^N \delta_1 (\hat{a}^\dagger_{n}\hat{a}_{n+1}+\hat{b}^\dagger_{n}\hat{b}_{n+1})-h.c.\nonumber\\
&&+\sum_n^N \delta_2 (\hat{a}^\dagger_{n}\hat{b}_{n+1}-\hat{a}^\dagger_{n}\hat{b}_{n-1})-h.c.,\label{eq:nonH_SSH}
\end{eqnarray}
with $a$ and $b$ the two sublattices of the system, and $N$ the total number of unit cells. 
Its Bloch Hamiltonian in momentum space reads
\begin{eqnarray}
h_{\rm 1D}(k) = (t_1+t_2\cos k)\sigma_x+t_2\sin k \sigma_y \nonumber \\    
+i(2\delta_1\sin k \sigma_0+2\delta_2\sin k\sigma_x),\label{eq:h_1D}
\end{eqnarray}
and the gap closing conditions are given by
\begin{eqnarray}
(t_1+t_2\cos k)^2+(t_2^2-4\delta^2)\sin^2 k&=&0,\nonumber\\
4\delta\sin k(t_1+t_2\cos k)&=&0.\label{eq:1D_gap_closing}
\end{eqnarray}
For $|2\delta_2|<|t_2|$, the above conditions are satisfied only 
at the DPs of the Hamiltonian, namely when $d_x=d_y=g_x=0$ at $(t_1,k)=(t_2,\pi)$ or $(-t_2,0)$.
In other words, conventional BBC is expected to be valid in this parameter regime, as shown by the PBC and OBC spectra in Fig. \ref{fig:1d_spectrum}(a) and (b).
The 1D topological properties can be characterized by the non-Hermitian Zak phase, defined as~\cite{kunst2018biorthogonal}
\begin{eqnarray}
  \gamma_{n} = -{\rm Im} \oint_{\rm BZ}  \langle \psi_{L,n}(k) | \partial_{k} | \psi_{R,n}(k) \rangle dk ,\nonumber 
\end{eqnarray}
with $\psi_{R,n}$ $(\psi_{L,n})$ the right (left)  biorthogonal~\cite{brody2014biorthogonal} Bloch eigenstate of the $n$th energy band.
In particular, in our considered two-band cases, the non-Hermitian Zak phase can be analytically expressed as a winding number of the real part of the Hamiltonian vector, as shown in Appendix~\ref{app:winding}.
As evident in Fig. \ref{fig:1d_spectrum}(c), 
$\gamma_n$ takes quantized values of $\pi$ and $0$ in the regimes with and without zero-mode edge states under OBC, respectively, reflecting the existence of conventional BBC. 
On the other hand, bulk states accumulate to the two ends of the 1D lattice under OBC, indicating the emergence of TC-NHSE, as shown in Fig. \ref{fig:1d_spectrum}(d). 
In addition, for the topological phase transition in Fig. \ref{fig:1d_spectrum}(d2), eigenstates are seen to be delocalized at the gap closing point, where the non-Hermitian terms of $\delta_1$ and $\delta_2$ vanish.

\begin{figure}[H]
  \includegraphics[width=1\linewidth]{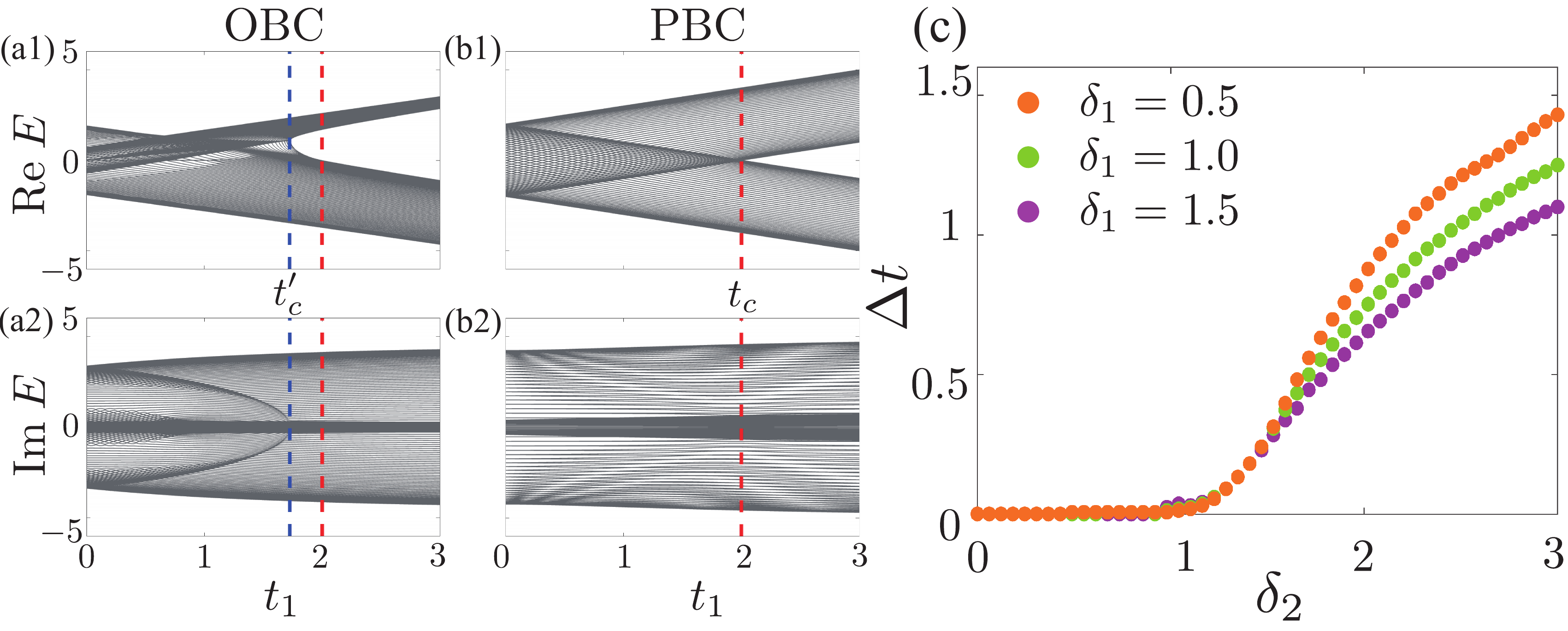}
  \caption{
    (a), (b) PBC and OBC eigenenergies versus $t_1$, with $t_{2}=1.5>t_2/2$ and $\delta_{1}=1$.
    The gap-closing points $t_{1}=t_{c}$ and $t_{1}=t_{c}'$ under the PBCs and OBCs are denoted by red and blue dash lines, respectively,
    which do not coincide.
    (c) The deviation between the PBC and OBC gap-closing points $\Delta t = t_{c}-t_{c}'$, as 
    a function of $\delta_{2}$ for three different values of $\delta_{1}$. 
    Other parameters : $t_{2}=2$ , $N=120$.
}
\label{fig:large delta2}
\end{figure}

When $|\delta_2|=|t_2/2|$, the system has two EPs at $\cos k=-t_1/t_2$, allowing the two bands to connect heads and tails with $|\delta_2|$ exceeding this critical value. Consequently, the conventional  BBC is broken for $|\delta_2|>|t_2/2|$, as shown in Fig. \ref{fig:large delta2}(a) and (b). On the other hand, $\delta_1$ does not enter the gap closing conditions of Eqs. \eqref{eq:1D_gap_closing}, thus it never breaks conventional BBC. In Fig. \ref{fig:large delta2}(c), we display numerical results of the deviation between PBC and OBC gap closing point as a function of $\delta_2$,
which acquires a non-vanishing value when $\delta_2> t_2/2$, regardless of the values of $\delta_1$.
Nonetheless, we point out that each of $\delta_1$ and $\delta_2$ alone can induce NHSE in this model,
as further demonstrated in Appendix~\ref{app:delta1_delta2}.
\bigskip



\section{TC-NHSE in semimetials}
Extending the 1D SSH model into 2D by introducing the same lattice structure along a second direction, we obtain a 2D Hamiltonian
\begin{eqnarray}
h_{\rm 2D}(k_x,k_y)&=&(t_1+t_2\cos k_x+t_2\cos k_y)\sigma_x\nonumber\\
&&+(t_2\sin k_x+t_2\sin k_y)\sigma_y.
\end{eqnarray}
This Hamiltonian generally describes a semimetal with two gapless points located along $k_x=-k_y$.
Following previous discussion, non-Hermiticity can be introduced in the same manners as for the 1D case to induce compatible NHSE, which does not affect these gapless points. 
To go beyond these two basic types of compatible NHSE, we take into account the spin degree of freedom, and consider two types of spin-orbit terms $h_{\rm SO}=h_{\rm SO,1}+h_{\rm SO,2}$,
\begin{eqnarray}
h_{\rm SO,1}&=&t_{\rm SO}(\sin k_x \tau_y-\sin k_y\tau_x),\\
h_{\rm SO,2}&=&v_{\rm SO}\sin (k_x+k_y+\phi)\tau_z\sigma_z,\nonumber
\end{eqnarray}
with $t_{\rm SO}$ and $v_{\rm SO}$ describing their strengths, and $\phi$ a phase factor to tune the position of Weyl points in momentum space.
Its eigenenergies read (subscribes ordered by their real energies)
\onecolumngrid
\begin{eqnarray}
E_{1,2,3,4}(k_{x},k_{y})=\pm\sqrt{
\begin{array}{c}
\left[\sqrt{(t_1+t_2\cos k_x+t_2\cos k_y)^2+(t_2\sin k_x+t_2\sin k_y)^2}\pm \sqrt{(t_{\rm SO}\sin k_x)^2+(t_{\rm SO}\sin k_y)^2}\right]^2 \\
+\left[v_{\rm SO}\sin (k_x+k_y+\phi)\right]^2\end{array}}~~.\nonumber \\ \label{eigenenergies_semi}
\end{eqnarray}
\twocolumngrid
The original gap closing points become four-fold degenerate Dirac points when the spin degree of freedom is introduced, 
which evolve into two-fold degenerate Weyl points at zero-energy under nonzero $v_{\rm SO}$ and $t_{\rm SO}$, whose location in the Brillouin zone is determined by
\begin{eqnarray}
{\rm Set~I:}~&&v_{\rm SO}\sin (k_x+k_y+\phi)=0,~\label{DP:con1}\\
&&(t_1+t_2\cos k_x+t_2\cos k_y)^2+(t_2\sin k_x+t_2\sin k_y)^2\nonumber\\
&&=(t_{\rm SO}\sin k_x)^2+(t_{\rm SO}\sin k_y)^2.\label{DP:con2}
\end{eqnarray}
Numerically, we find that these Weyl points can be locally gapped out only by terms anticommuting with $h_{\rm SO,2}$, namely $\sigma_{\alpha}\tau_{\bar{\alpha}}$ with $\alpha,\bar{\alpha}\in\{x,y,z\}$, $\alpha\neq\bar{\alpha}$, and either $\alpha$ or $\bar{\alpha}$ being $z$.
Thus the existence of the zero-energy Weyl points can be attributed to the protection of an parity-time (PT) symmetry, 
\begin{eqnarray}
\sigma_x\tau_xh^*(k)\sigma_x\tau_x=h(k), \label{eq:symmetry1}
\end{eqnarray}
which excludes the above mentioned matrices from the Hamiltonian.

\begin{figure}
  \includegraphics[width=1\linewidth]{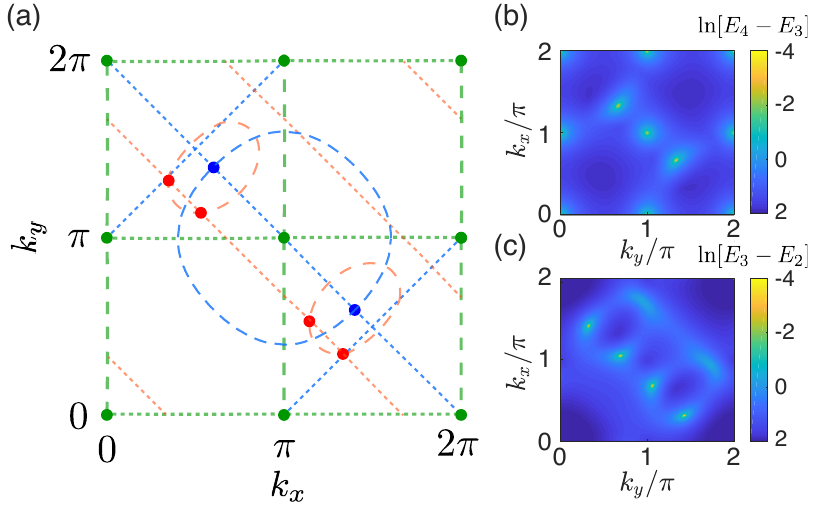}
  \caption{(a)
      Schematic of the position (points) and degenerate conditions (dash lines) 
      of Set I (red), II (blue) and III (green) Weyl points in momentum space. 
      (b) energy difference between $E_3$ and $E_4$, zeros indicate the Weyl points of sets II and III [blue and green points in (a)].
      (c) energy difference between $E_2$ and $E_3$, zeros indicate the Weyl points of set I [red points in (a)].
      Parameters for (b) and (c) are
      $t_1=t_2=t_{\rm SO}=2$, $v_{\rm SO}=1.2$, and $\phi=\pi/4$.
        }
  \label{fig:DP_2D_sketch}
\end{figure}

Other than these zero-energy gap closing points, extra Weyl points also emerge between the first (last) two bands 
when
\begin{eqnarray}
{\rm Set~II:}~t_1+t_2\cos k_x+t_2\cos k_y\nonumber\\=t_2\sin k_x+t_2\sin k_y=0,\nonumber
\end{eqnarray}
or
\begin{eqnarray}
{\rm Set~III:}~t_{\rm SO}\sin k_x=t_{\rm SO}\sin k_y=0.\nonumber
\end{eqnarray}
Note that these two sets of Weyl points are determined solely by $h_{\rm 2D}$ and $h_{\rm SO,1}$, respectively.
In order words, Weyl points of Set II and III are protected by hidden PT symmetries of the corresponding sub-Hamiltonians,
\begin{eqnarray}
\sigma_x h^*_{\rm 2D}(k_x,k_y)\sigma_x=h_{\rm 2D}(k_x,k_y),\label{eq:symmetry2}\\
\tau_x h^*_{\rm SO,1}(k_x,k_y)\tau_x=h_{\rm SO,1}(k_x,k_y),\label{eq:symmetry3}
\end{eqnarray}
 respectively.
The locations in the Brillouin zone of these three sets of Weyl points are sketched in Fig. \ref{fig:DP_2D_sketch}(a). 
In Fig. \ref{fig:DP_2D_sketch}(b) and (c), we demonstrate the energy difference between the highest and middle two bands, respectively,
where zeros of the former (later) represents Weyl points of Sets II or III (Set I).

\begin{figure}
  \includegraphics[width=1\linewidth]{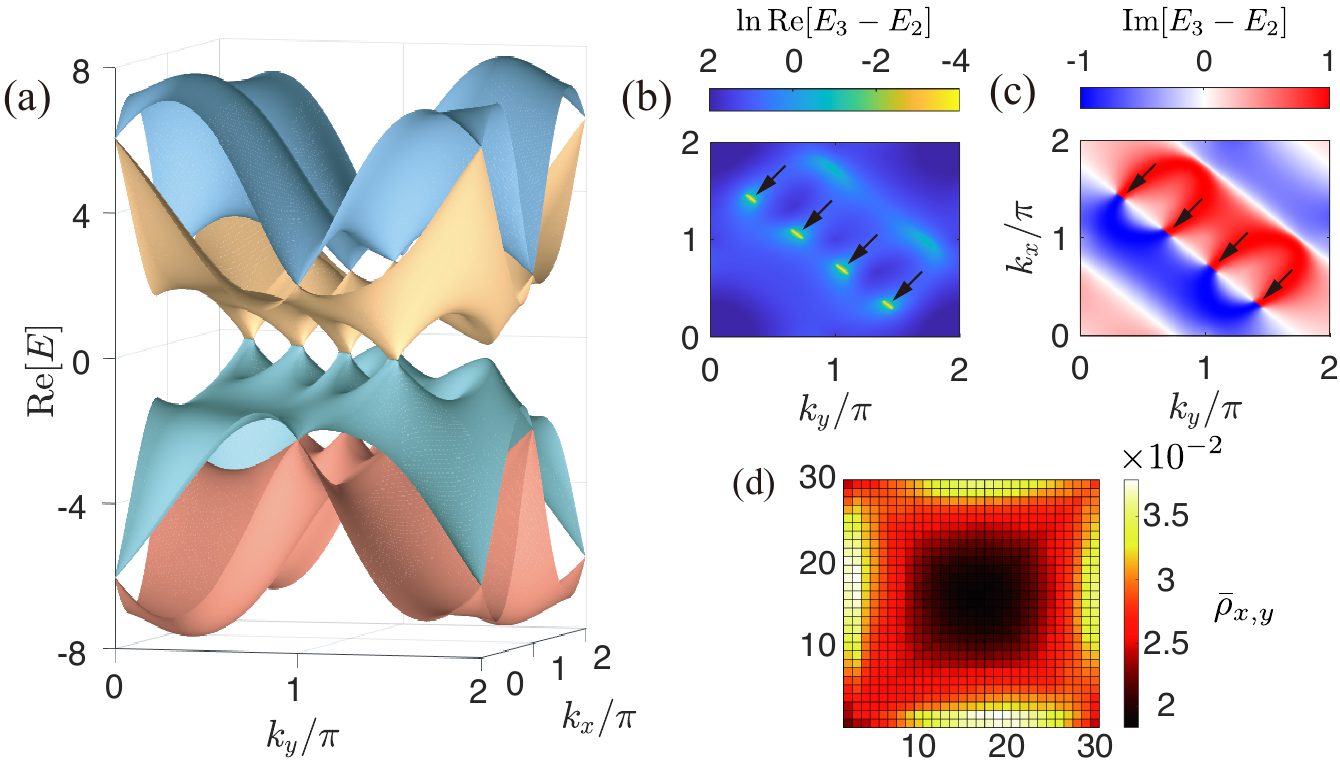}
  \caption{  
  The properties of the system when introducing the non-Hermitian term $h_{{\rm nonH}1}$ with $\delta_{1}=0.5$.
  (a) The real part of the energy bands structure, valence and conduction bands meet at bulk Fermi arcs. 
  The logarithm of real energy difference (b) and the imaginary energy difference (c) between the middle two bands $E_{2}$ and $E_{3}$.
  Black arrows indicate the regions of bulk Fermi arcs, where the imaginary energies switch between the two bands, as shown by the jump of ${\rm Im} [E_3-E_2]$ between large negative and positive values across the bulk Fermi arcs.
  (d) The average distribution $\bar{\rho}_{x,y}$ of all eigenmodes under OBCs. 
  Other parameters: $t_1=t_2=t_{\rm SO}=2$, $v_{\rm SO}=1.2$, and $\phi=\pi/4$.
  }
  \label{fig:semi2d_1}
 \end{figure}
Given the coexistence of Weyl points between different bands, we argue that NHSE can not only be introduced in a way that is compatible to all sets of Weyl points, e.g., through introducing a non-Hermitian $\tau_0$ term, but also in a way that is selectively compatible to different sets of Weyl points,
turning some of them into EPs while the others remain unchanged.
As a first demonstration, we consider 
the following non-Hermitian term as an example,
  \begin{eqnarray}
    h_{{\rm nonH}1} = i\delta_1\tau_{z}\sigma_{z},
  \end{eqnarray}  
which breaks the PT symmetry and turns the first set of Weyl points (at ${\rm Re}[E]=0$) into bulk Fermi arcs~\cite{kozii2017nonhermitian,pujol-closa2022dirac,observation} with degenerate real energies, as shown in Fig. \ref{fig:semi2d_1}(a) and (b). In contrast to boundary Fermi arcs in Hermitian systems, 
bulk Fermi arcs are terminated at EPs with zero imaginary energy, 
whose locations are determined by
 \begin{eqnarray}
  k_{x}+k_{y}+\phi&=&0,\nonumber\\
  |E_{2,3}(k_{x},-k_{x}-\phi)| &=& |\delta_{1}|,\nonumber
\end{eqnarray}
and the two bands sorted by their real energies have discontinuous imaginary energies across the arcs, as shown in Fig. \ref{fig:semi2d_1}(c).

On the other hand, 
$h_{{\rm nonH}1}$ does not alter the hidden symmetries of the sub-Hamiltonians, therefore
the other two sets of Weyl points remains unchanged, as can be seen in Fig. \ref{fig:semi2d_1}(a) between the top two or the bottom two bands.
In Fig. \ref{fig:semi2d_1}(d), we display the average distribution of all eigenmodes under OBCs,
$$\bar{\rho}_{x,y}=(1/N)\sum\limits_{i}^{N}\sqrt{\sum\limits_{\alpha} |\psi^{i}_{x,y,\bm{\alpha}}|^2}, $$
with $N$ the size of the Hilbert space, and $\psi^{i}_{x,y,\alpha}$ the wave amplitude of the $i$th eigenmode at unit cell $(x,y)$, $\bm{\alpha}=(\tau,\sigma)$ denoting the spin and sublattice degrees of freedom respectively. 
A clear edge localization can be seen in the figure, indicating the emergence of NHSE compatible with the second and third sets of topological Weyl points.
 
\begin{figure}
  \includegraphics[width=1\linewidth]{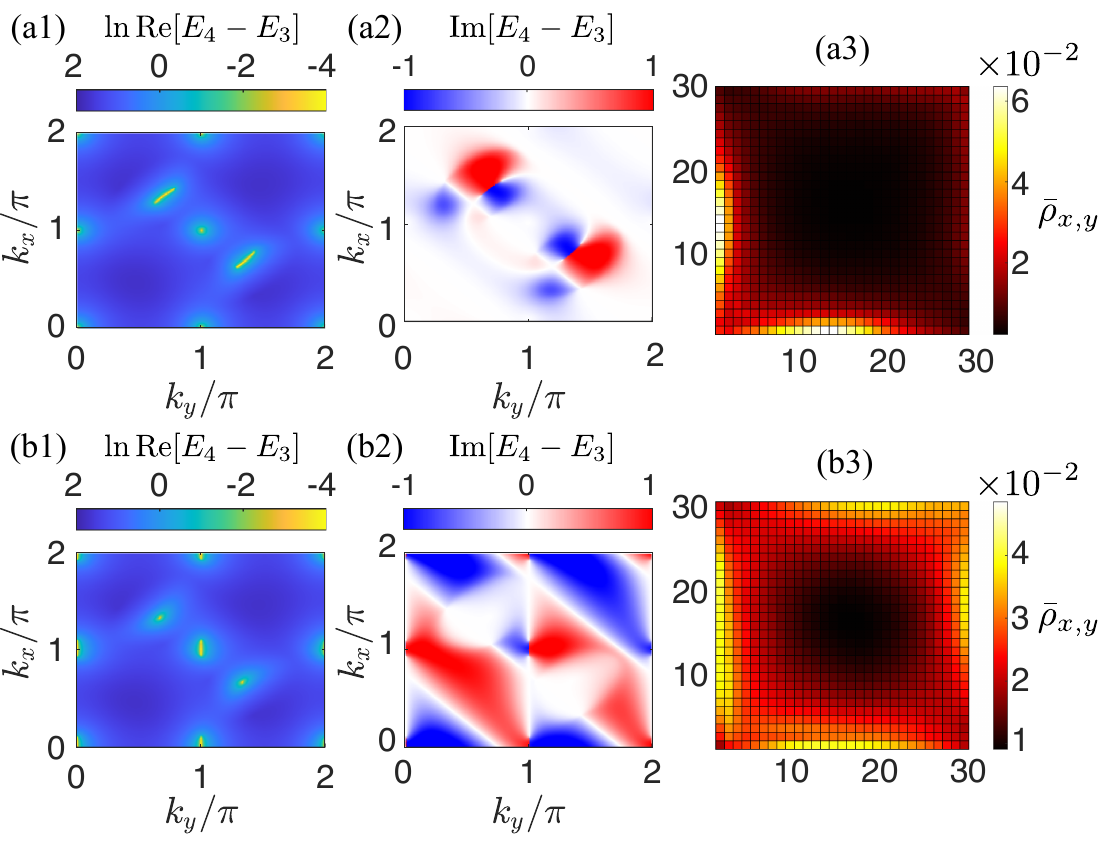}
  \caption{
    The properties of the system when introducing the non-Hermitian term $h_{{\rm nonH}2}$ (a) or $h_{{\rm nonH}3}$ (b) with $\delta_{2}=0.35$ or $\delta_{3}=0.35$.
    The logarithm of real energy difference (a1, b1) and the imaginary energy difference (a2, b2) between the last two bands $E_{3}$ and $E_{4}$.
    (a3, b3) The average distribution $\bar{\rho}_{x,y}$ of all eigenmodes under OBCs. 
    Other parameters: $t_1=t_2=t_{\rm SO}=2$, $v_{\rm SO}=1.2$, $\phi=\pi/4$.
  }
  \label{fig:delta23}
\end{figure}  

\begin{figure*}
  \includegraphics[width=0.6\linewidth]{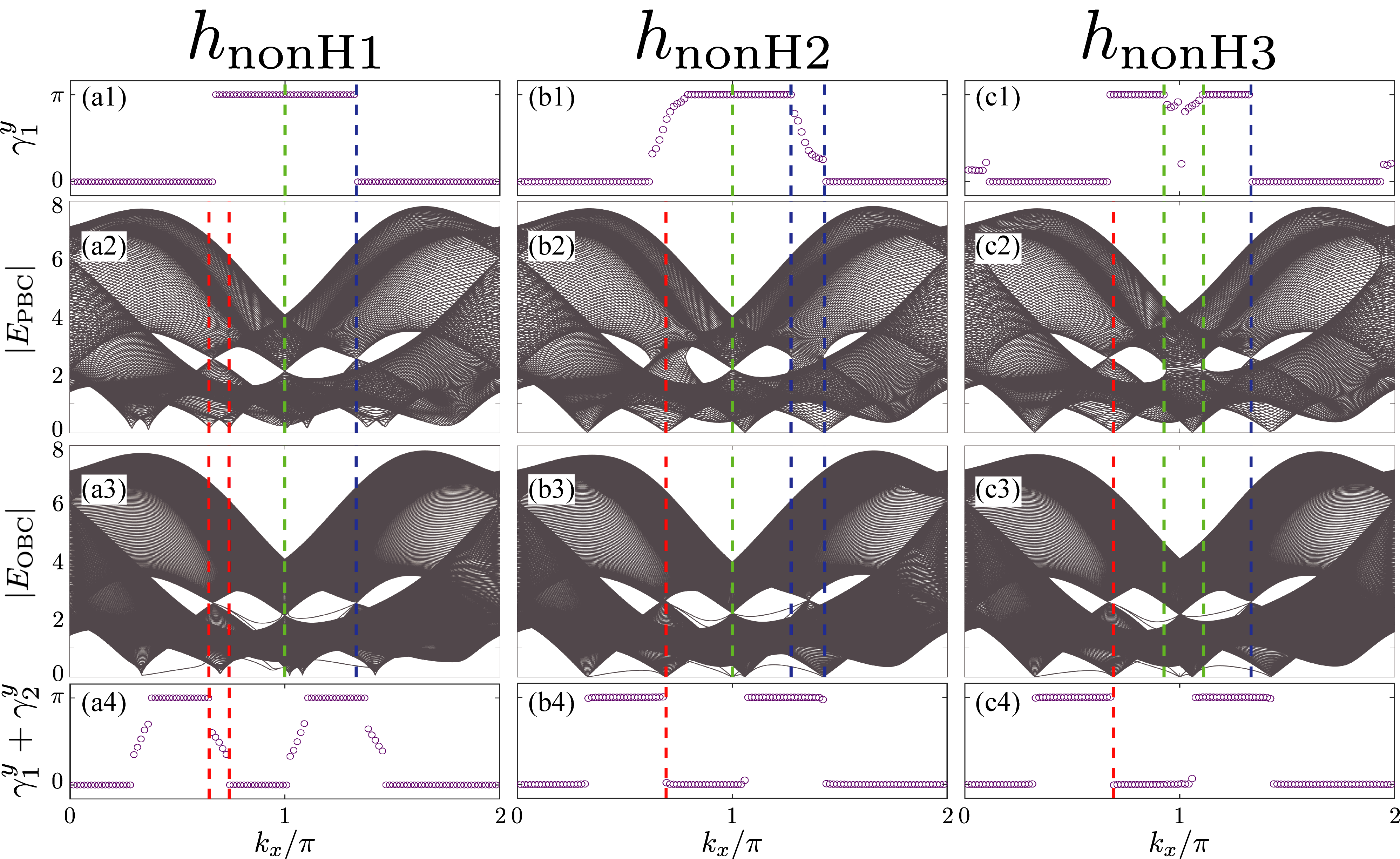}
  \caption{
      The absolute value of $x$-PBC/$y$-PBC (a$2$, b$2$, c$2$) and $x$-PBC/$y$-OBC (a$3$, b$3$, c$3$) spectrum as a function of $k_{x}$ with introducing the non-Hermitian terms $h_{{\rm nonH}1}$ (a), $h_{{\rm nonH}2}$ (b) and $h_{{\rm nonH}3}$ (c)
      into the Hermitian Hamiltonian $h_{\rm 2D}+h_{\rm SO}$, corresponding to the change of Zak phase $\gamma^{y}_{1}$ (a$1$, b$1$, c$1$) and $\gamma^{y}_{1}+\gamma^{y}_{2}$ (a$4$, b$4$, c$4$). 
      The red (Set I), blue (Set II) and green (Set III) dash lines indicate the gap-closing points under the PBC in $y$-direction 
      around one of each set of Weyl points. 
      It shows that different sets of Weyl points are turned into EPs by differnent non-Hermitian terms and the Zak phase isn't quantized when EPs appear.  
      The length in $y$-direction is $N=150$ and the non-Hermitian parameters are $\delta_{1}=0.5$ (a), $\delta_{2}=0.35$ (b) and $\delta_{3}=0.35$ (c) respectively.
      Other parameters: $t_{1}=t_{2}=t_{\rm SO}=2,v_{\rm SO}=1.2,\phi=\pi/4$.
  }
  \label{fig:2D_Berry}
\end{figure*}

Similarly, by breaking different hidden PT symmetries of Eqs. \eqref{eq:symmetry2} and \eqref{eq:symmetry3}, the resultant non-Hermitian system may host NHSE (see Fig. \ref{fig:delta23} (a3, b3)) compatible with Set I and only one of the rest two sets of Weyl points. 
Specifically, we consider the following two non-Hermitian terms,
  \begin{eqnarray}
    h_{{\rm nonH}2} &=& i2\delta_2 \sin(k_{x}+k_{y}+\phi) \sigma_{x},   \\
    h_{{\rm nonH}3} &=& i2\delta_3 \sin(k_{x}+k_{y}+\phi) \tau_{x}.
  \end{eqnarray}
Note the these extra terms also inevitably break the PT symmetry of Eq. \eqref{eq:symmetry1} for the overall system. 
To avoid destroying the zero-energy Weyl points of Set I,
The explicit forms of $h_{{\rm nonH}2}$ and $h_{{\rm nonH}3}$ are chosen to vanish (thus recovering the symmetry) at the momenta of these points.
  In Fig. \ref{fig:delta23}(a1, a2) [(b1, b2)], the degenerate real energies and discontinuous imaginary energies show that the second (third) set of Weyl points
  are turned into bulk Fermi arcs due to $h_{{\rm nonH}2}$ ($h_{{\rm nonH}3}$), which breaks the corresponding hidden PT symmetry, similar to the influence of $h_{{\rm nonH}1}$ on the first set of Weyl points.
Different types of NHSE are also seen to emerge in both cases, as shown in Fig. \ref{fig:delta23}(a3) and (b3),
which are topologically compatible with the other two sets of Weyl points that remain unchanged as in the Hermitian case (sets I and III for  $h_{{\rm nonH}2}$, and sets I and II for  $h_{{\rm nonH}3}$).
We also note that extra EPs and bulk Fermi arcs (irrelevant to the Weyl points) may appear between different bands when further increasing each of the non-Hermitian parameters $\delta_{1,2,3}$ (no shown).

To further demonstrate the topological compatibility of NHSE induced by different non-Hermitian terms, we compare the absolute values of eigen-spectra under full PBC and $x$-PBC/$y$-OBC in Fig. \ref{fig:2D_Berry}. It is seen that in the presence of each non-Hermitian term, 
only one set of Weyl points are turned into EPs under full PBC, where the gap closing points diverge under different boundary conditions.
For example, 
in Fig. \ref{fig:2D_Berry}(a) with nonzero $\delta_1$, 
the gap closing points at $|E|=0$ do not coincide under PBC (partly marked by red dash lines) and OBC,
in contrast to those between the two bands of $|E|$ (sets II and III, partly marked by blue and green dash lines respectively).

Taking $k_x$ as a parameter, topological properties related to these gap closing points can be characterized by the Zak phase as in the 1D case, defined as
\begin{eqnarray}
  \gamma_{n}^{y}(k_{x}) = - {\rm Im} \oint_{0}^{2\pi} dk_{y} \langle \psi_{L,n}(k_{x},k_{y}) | \partial_{k_{y}} | \psi_{R,n}(k_{x},k_{y}).\rangle \nonumber
\end{eqnarray} 
As seen in Fig. \ref{fig:2D_Berry}(a), $\gamma_1^y$ (which describes the topological properties of the lowest energy band) is always quantized at $\pi$ or $0$, and jumps between them only at the gap closing points (Weyl points) between the two bands of $|E|$.
Note that it remains unchanged across the gap closing point at $k_x=\pi$, which is actually the projection of a pair of Weyl points with opposite topological charges (at $k_y=0$ and $\pi$).
On the other hand, the summed Zak phase between the lower two bands, $\gamma_1^y+\gamma_2^y$, becomes non-quantized  in the parameter regime between each pair of EPs at $|E|=0$, 
and its transition points do not match the gap closing points under OBC. 
Similarly, nonzero $\delta_2$ and $\delta_3$ lead to a non-quantized Zak phase $\gamma_1^y$ in the parameter regimes between the corresponding EPs induced by them respectively, while $\gamma_1^y+\gamma_2^y$ is always quantized and only jumps discontinuously at $|E|=0$, as shown in Fig. \ref{fig:2D_Berry}(b) and (c).
These observations further 
validate the topological compatibility of NHSE induced by different non-Hermiticity in our model with different sets of Weyl points.

\section{Summary}
To summarize, we have revealed a class of TC-NHSE, which induces non-trivial spectral winding and skin accumulation of eigenstates to the system without breaking conventional BBC. 
 By analyzing the degenerate condition of general non-Hermitian two-band Hamiltonians, we have provided a general scheme for introducing NHSE without inducing any EP, and demonstrate the corresponding TC-NHSE in a 1D non-Hermitian SSH model.
Specifically, the induced NHSE is compatible with the system's topology either in a universal manner, or only in a certain parameter regime, depending on how the non-Hermiticity enters the Hamiltonian explicitly.
  Next, we extended the 1D SSH model into a 2D $\mathcal{PT}$-symmetric semimetal with three sets of Weyl points between different energy bands, protected by different symmetries or hidden symmetries of the system respectively.
  We find that by introducing non-Hermitian terms that break different symmetries of the system,
  the corresponding TC-NHSE can be selectively compatible to one set of Weyl points (i.e. remain unchanged in the momentum space) and transform the rest into pairs of EPs connected by bulk Fermi arcs.
By taking $k_{x}$ as a parameter, these unaffected Weyl points can be characterized by the quantized non-Hermitian Zak phase, which takes non-quantized values in the parameter regime between each pair of EPs transformed from a Weyl point,
  reflecting the selective compaticity of the TC-NHSE in the system.

The topological properties of our studied systems are characterized by the non-Hermitian Zak phases, which essentially describes 1D topology corresponding to edge states gapped from the bulk (as in 1D topological insulators), or bulk degeneracy with co-dimension of two (as in 2D semimetals). Nevertheless,
our scheme for inducing TC-NHSE relies only on the vanishing of non-Hermiticity at bulk degeneracy,
and thus it can also be straightforwardly extended to other higher-dimensional topological phases, such as 2D topological insulators characterized by the Chern number.
A potential area of interest is the 
generalization of TC-NHSE in more sophisticated topological phases, such as higher-order topological phases where topological edge states may not directly correspond to bulk topology~\cite{benalcazar2017quantized,song2017dimensional,jia2023unified,khalaf2018higherorder,lei2022topological,luo2019higherorder}, or hybrid skin-topological phases where NHSE interplay with topological localization in a distinctive manner~\cite{lee2019hybrid,li2022gainlossinduced,zhu2022hybrid}.

\section{Acknowledgement}
This work is
supported by National Natural Science Foundation of China (Grant No. 12104519) and the
Guangdong Project (Grant No. 2021QN02X073).

\appendix

\section{The non-Hermitian Zak phase expressed as a winding number for chiral non-Hermitian system}\label{app:winding}

Here we consider a general non-Hermitian two-band Hamiltonian in momentum space which can be written in the form of Eq. (\ref{eq:2band}),
\begin{eqnarray}
h(k)=\sum_{\alpha=0,x,y,z}h_\alpha(k)\sigma_\alpha,\nonumber\\
h_{\alpha}=d_{\alpha}(k)+ig_{\alpha}(k).
\end{eqnarray}
Its left and right eigenstates are given by 
\begin{eqnarray}
  |\psi_+^{R}\rangle = 
  \begin{pmatrix}
    e^{-i\phi}\cos\frac{\theta}{2} \\
    \sin\frac{\theta}{2}
  \end{pmatrix},\quad
  |\psi_-^{R}\rangle = 
  \begin{pmatrix}
    e^{-i\phi}\sin\frac{\theta}{2} \\
    -\cos\frac{\theta}{2}
  \end{pmatrix},
\end{eqnarray} 
and
\begin{eqnarray}
  \langle\psi_+^{L}| = 
  \begin{pmatrix}
    e^{i\phi}\cos\frac{\theta}{2} \\
    \sin\frac{\theta}{2}
  \end{pmatrix},\quad
  \langle\psi_-^{L}| = 
  \begin{pmatrix}
    e^{i\phi}\sin\frac{\theta}{2} \\
    -\cos\frac{\theta}{2}
  \end{pmatrix},
\end{eqnarray}
with $\cos\theta=h_z/\sqrt{h_x^2+h_y^2+h_z^2}$ and $\cos\phi=h_x/\sqrt{h_x^2+h_y^2}$, and $\pm$ denoting the two bands.
We note that $h_0$ does not enters the expressions of eigenstates, and we will omit this term in our following discussion.
Here we consider the non-Hermitian Zak phase of the lower band, which can be obtained as \cite{teo2020topological}
\begin{eqnarray}
  \gamma^{(-)} &= &
  -{\rm Im} \oint_{\rm BZ}  \langle \psi^{L}_{-} | \partial_{k} | \psi^{R}_{-} \rangle dk \nonumber \\
  &=& {\rm Re}\oint_{\rm BZ}\frac{\partial\phi}{\partial k}\sin^2\frac{\theta}{2}dk.
\end{eqnarray} 
If the Hamiltonian has chiral symmetry $\sigma_zh(k)\sigma_z=-h(k)$, we have $\cos\theta=0$,
and $\gamma^{(-)}$ is given ${\rm Re}\oint_{\rm BZ}\frac{d\phi}{2}$, namely half the total winding angle of $(h_x,h_y)$ when $k$ varies from $0$ to $2\pi$, or in other words, the winding number of $(h_x,h_y)$ multiplied by $\pi$. For non-Hermitian systems, $\phi$ may take complex values, making the winding number of $(h_x,h_y)$ ill-defined. Thus we further consider the winding number of ($d_x$,$d_y$), i.e. the Hermitian parts of a chiral-symmetric Hamiltonian,
and their winding number $\nu$ in the parameter space takes the form as 
\begin{eqnarray}
  \nu &=& \frac{1}{2\pi} \oint_{\mathcal{C}} d\phi' \\
  &=& \frac{1}{2\pi} \int_0^{2\pi} dk \frac{d_x\partial_kd_y-d_y\partial_kd_x}{d_x^2+d_y^2},
\end{eqnarray} 
where $\mathcal{C}$ is the trajectory of ($d_x$,$d_y$) in a period.
Further considering the non-Hermitian terms $g_x$ and $g_y$ vanish together with the Hermitian parts of the Hamiltonian (so to generate TC-NHSE):
\begin{eqnarray}
  g_x=ad_y,\quad g_y=bd_x
\end{eqnarray} 
where $a$ and $b$ are chosen to be real and positive without loss of generality, the Zak phase can be expressed as
\begin{eqnarray}
  \gamma^{(-)} = \frac{\rm Re}{2} \oint_{\rm BZ} \frac{(1+ab)(d_x\partial_kd_y-d_y\partial_kd_x)}{(1-b^2)d_x^2+(1-a^2)d_y^2+2i(a+b)d_xd_y}.  \nonumber \\
  \label{eq:zak}
\end{eqnarray} 
Substituting $d\phi'=dk  (d_x\partial_kd_y-d_y\partial_kd_x)/(d_x^2+d_y^2) $ into Eq. (\ref{eq:zak}), we obtain
\onecolumngrid
\begin{eqnarray}
  \gamma^{(-)} = \frac{\rm Re}{2} \int_{\phi'(k=0)}^{\phi'(k=2\pi)} \frac{(1+ab)d\phi'}{(1-b^2)\cos^2\phi'+(1-a^2)\sin^2\phi'+2i(a+b)\sin\phi'\cos\phi'},   \label{eq:zak2}
\end{eqnarray}
\twocolumngrid
which is nonzero only when the winding of $\phi'$ is nontrivial (i.e. $\nu\neq0$), as $\phi'(0)=\phi'(2\pi)$.
To further simplify the above integral, we substitute $z=\tan\phi'$ into Eq. (\ref{eq:zak2}) and obtain
\begin{eqnarray}
  \gamma^{(-)}  &=&\frac{1}{2}{\rm Re} \int dz \frac{1+ab}{(1-b^2)+(1-a^2)z^2+2i(a+b)z} \nonumber \\
   &=& {\rm Re} \frac{1+ab}{2(1-a^2)} \int dz  \frac{1}{(z-z_1)(z-z_2)}. \nonumber 
\end{eqnarray}
Here $z_1$ and $z_2$ are the two roots of the denominator polynomial, which take pure imaginary values, given by
\begin{eqnarray}
  z_{1,2} = i \frac{-(a+b)\pm(ab+1)}{1-a^2}.
\end{eqnarray}
In nontrivial cases with $\nu\neq 0$, $\phi'$ varies in the interval $\left[ \phi_{0}, \phi_{0}+2\nu\pi \right]$, and $z$ varies $\nu$ times in infinite intervals $\left[ \tan\phi_{0}, +\infty \right)$ , $\left( -\infty, +\infty \right)$ and $\left(-\infty, \tan\phi_{0} \right]$, so
\begin{eqnarray}
  \gamma^{(-)} &=& {\rm Re} \frac{1+ab}{2(1-a^2)}  \int_{\phi=\phi_0}^{\phi=\phi_0+2\nu\pi} dz \frac{1}{(z-z_1)(z-z_2)}  \nonumber  \\
               &=&  {\rm Re} \frac{1+ab}{1-a^2}  \oint_{\mathcal{L}+\mathcal{C}'}  \frac{\nu}{(z-z_1)(z-z_2)}  \label{eq:zak3}
\end{eqnarray}
where $\mathcal{L}$ is the real axis and $\mathcal{C}'$ is a semicircle of infinite radius on the lower complex plane. 
The integral of Eq. (\ref{eq:zak3}) can be calculated by residue theorem
\begin{eqnarray}
  \gamma^{(-)} &=& {\rm Re} \frac{1+ab}{1-a^2} \cdot 2\pi i \sum\limits_{j} {\rm Res} \left[ \frac{\nu}{(z-z_1)(z-z_2)} \right]  \nonumber \\
               &=& 
               \begin{cases}
                -\nu\pi, \quad a,b>1 \\
                \nu\pi,  \quad a,b<1 \\
                0, \quad  a>1,b<1 \quad {\rm or} \quad a<1,b>1
               \end{cases}  \label{eq:zak4} \\
\end{eqnarray}
where $z_{j}$ is the residue of the function $\frac{1}{(z-z_1)(z-z_2)}$ on the lower complex plane. 
For the non-Hermitian SSH model we discussed in Sec. \ref{sec:1d}, the parameters are given by $a=0$ and $b=2\delta_{2}/t_{2}$.
We can see that the Zak phase can be mapped to the winding number $\nu$ when $\delta_{2}<t_2/2$, i.e. the parameter regime with TC-NHSE where the conventional BBC is valid.
However, when $\delta_2>t_2/2$, the Zak phase $\gamma^{(-)}$ is always $0$, and can no longer predict the existence of zero-mode edge states.

\section{TC-NHSE in the 1D non-Hermitian SSH model}\label{app:delta1_delta2}

\begin{figure}
  \includegraphics[width=1\linewidth]{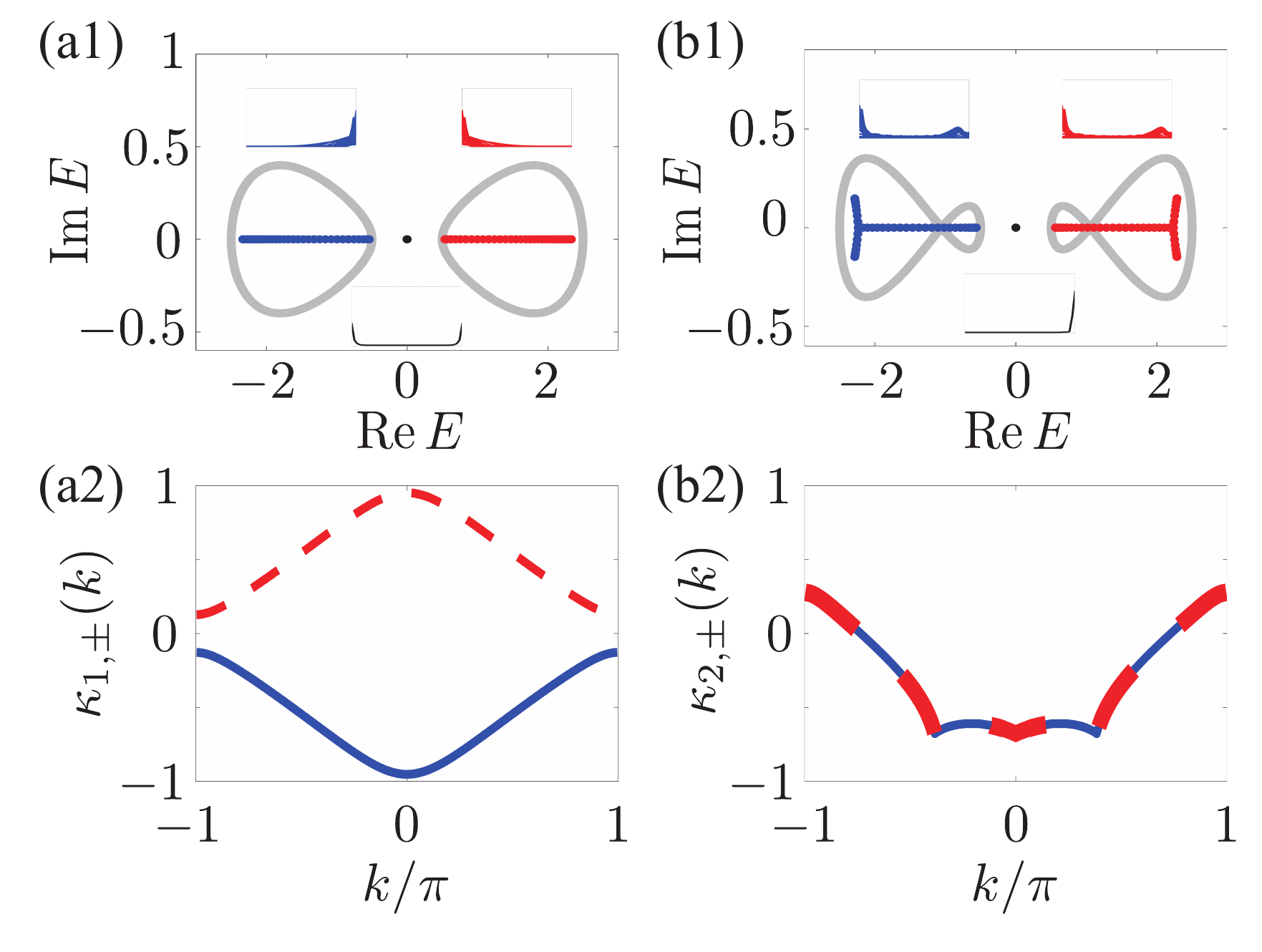}
  \caption{
  (a1) and (b1) complex spectra under the PBCs (gray) and OBCs (red, blue, and black) for (a) $\delta_1=0.2$, $\delta_2=0$ and (b) $\delta_1=0$, $\delta_2=0.25$, respectively, with insets demonstrating distribution profiles of OBC eigenstates with the same colors.
  (a2) and (b2) The inverse localization length $\kappa_{\pm}(k)$ for the two bands marked by the same colors in (a1) and (b1). 
  In (b2), blue and red curves overlap as $\kappa_{2,+}(k)=\kappa_{2,-}(k)$.
  Other parameters are $t_1=1$, $t_{2}=1.5$ and $N=40$. 
  }
  \label{fig:appB}
\end{figure}

In this Appendix, 
we provide further analysis of the NHSE induced by $\delta_1$ and $\delta_2$ in the model described by Eq. (\ref{eq:h_1D}).
As shown in Fig. \ref{fig:appB}(a1) and (b1), each of these two non-Hermitian parameters along can induce NHSE in the system, yet the resultant skin modes show different localizing directions. Therefore we shall treat these two parameters separately in the following discussion.

We first give a systematic analysis of the case in the presence of a nonzero $\delta_1$.
The Bloch Hamiltonian becomes
\begin{eqnarray}
h_{1}(k)=(t_1+t_2\cos k)\sigma_x+t_2\sin k \sigma_y+i2\delta_1\sin k \sigma_0, \nonumber 
\end{eqnarray}
with its eigenenergies given by
\begin{eqnarray}
E_{1,\pm}(k) = i2\delta_1\sin k\pm\sqrt{t_1^2+t_2^2+2t_1t_2\cos k},\label{eq:E_delta1}
\end{eqnarray}
supporting the same DP as the Hermitian SSH model.
Moreover, $h_1(k)$ satisfy
\begin{eqnarray}
\sigma_{y}h_1(k)\sigma_{y} = -h_1^{*}(k),
\end{eqnarray}
so that the eigenenergies of $h_1(k)$ are always symmetric about the imaginary axis (${\rm Re}E=0$), i.e. $E_+(k)=-E_-^*(k)$, as shown in Fig \ref{fig:appB} (a1).
Therefore a real line-gap always presences between the two bands $E_{1,\pm}$ except when they touch at ${\rm Re}E=0$, avoiding the band overlapping problem.

In 1D systems, NHSE can be quantitatively described by 
generalizing the Bloch Hamiltonian to a non-Bloch one, $H(k)\rightarrow H(k+i\kappa(k))$,
where $\kappa(k)$ describes the inverse localization length of skin modes~\cite{yao2018edge,lee2019anatomy}.
The eigenenergies of Eq. (\ref{eq:E_delta1}) satisfy
\begin{eqnarray}
  E_{1,+}(k+i\kappa) = -E_{1,-}^{*}(k-i\kappa), \label{eq:d1_con1}
\end{eqnarray}
where $\kappa$ is a arbitrary constant inverse localization length.
Assuming the inverse localization length is given by $\kappa_{\pm}(k)$ for two bands,
the OBC eigenenergies must have paired $k$ and $k'$ satisfying
\begin{eqnarray}
  E_{\pm} \left[ k+i\kappa_{\pm}(k) \right] = E_{\pm} \left[ k'+i\kappa_{\pm}(k)\right]. \label{eq:OBC_kappa}
\end{eqnarray}
The exact relation between $k$ and $k'$ is unknown, yet from Eqs. (\ref{eq:d1_con1}) and (\ref{eq:OBC_kappa}) we will also have
\begin{eqnarray}
  E_{1,-}\left[ k-i\kappa_{1,+}(k) \right] = E_{1,-} \left[ k'-i\kappa_{1,+}(k) \right]. \label{eq:d1_con2}
\end{eqnarray}
Combining the Eqs. (\ref{eq:OBC_kappa}) and (\ref{eq:d1_con2}), we reach the conclusion that 
\begin{eqnarray}
  \kappa_{1,+}(k) = -\kappa_{1,-}(k),
\end{eqnarray}
meaning that the eigenstates have opposite localizing directions for the two bands, as shown in the Fig. \ref{fig:appB} (a2) and the insets of Fig. \ref{fig:appB} (a1).

Similar analytics can be applied to the case in the presence of a nonzero $\delta_2$.
The Hamiltonian in momentum space is given by  
\begin{eqnarray}
  h_{2}(k) = (t_1+t_2\cos k+i2\delta_2\sin k)\sigma_x+t_2\sin k \sigma_y,  \nonumber  
\end{eqnarray} 
which satisfies the chiral symmetry
\begin{eqnarray}
\sigma_zh_2(k)\sigma_z=-h_2(k).
\end{eqnarray}
The eigenvalues of $h_2(k)$ appear in $(E,-E)$ pairs,
\begin{eqnarray}
 E_{2,\pm}(k) = \pm\sqrt{(t_1+t_2\cos k+i2\delta_2\sin k)^2+(t_2\sin k)^2},\label{eq:E_delta2} \nonumber \\
\end{eqnarray}
as shown in Fig \ref{fig:appB} (b1). 
With the non-Bloch generalization taken into account,
 the eigenenergies satisfies
\begin{eqnarray}
  E_{2,+}(k+i\kappa) = -E_{2,-}(k+i\kappa).\label{eq:d2_con}
\end{eqnarray}
Combining Eqs. (\ref{eq:OBC_kappa}) and (\ref{eq:d2_con}) we shall obtain
\begin{eqnarray}
  \kappa_{2,+}(k) = \kappa_{2,-}(k),
\end{eqnarray}
meaning that the eigenstates have the same localizing direction for the two bands, as shown in the Fig. \ref{fig:appB} (b2) and the insets of Fig. \ref{fig:appB} (b1).

\bibliography{references.bib}

\end{document}